\documentstyle[11pt,paspconf,epsf]{article}

\begin{document}

\title{Gravitational Lensing of High Redshift Sources}

\author{Rennan Barkana and David W. Hogg}
\affil{Institute for Advanced Study, Olden Lane, Princeton,
NJ 08540}

\author{Abraham Loeb}
\affil{Astronomy Department, Harvard University, 60 Garden 
St., Cambridge, MA 02138}

\author{Roger Blandford}
\affil{130-33 California Institute of Technology, Pasadena, CA 91125}

\begin{abstract}
The combination of deep exposures and high resolution offered by
telescopes in space allows the detection of lensing over a wide range
of source redshifts and lens masses. As an example, we model a lens
candidate found in the southern Hubble Deep Field. The system consists
of a source galaxy lensed into an arc, $0\farcs9$ from an elliptical
galaxy. The photometric redshift of 0.6 for the lens galaxy implies a
mass-to-light ratio of 15 in solar units, out to three effective
radii. This lens system may be a preview of the large number of lensed
high-redshift galaxies that will be detected with the Next Generation
Space Telescope ({\it NGST}\,). When {\it NGST}\, is launched in the
next decade, some of the earliest galaxies and quasars in the Universe
may be observed. Popular models of structure formation imply that at a
given observed flux, roughly $3\%$ of redshift $>5$ sources are multiply
imaged. Thus, {\it NGST}\, should detect several lensed objects in
each field of view. In addition, {\it NGST}\, will be a valuable tool
for weak lensing measurements, as long as it can resolve the
background galaxies. We estimate the angular size distribution of high
redshift sources within hierarchical models of structure formation and
find that most will be resolved by {\it NGST}\, even at $z>10$.
\end{abstract}


\section{Introduction}

As detailed throughout these proceedings, gravitational lensing
studies have yielded a wealth of information on galaxies. As
technological advances in astronomy lead to the discovery of new
populations of sources, the techniques of strong and weak
gravitational lensing can be applied in order to study these source
populations and their lenses.

The extraordinary resolution of the Hubble Space Telescope ({\it
HST}\,) has allowed for the first time the detection of lenses where
both the source and the deflector are ``normal'' optically-selected
galaxies. Ratnatunga et al.\ (1995) discovered two such lenses. The
Southern Hubble Deep Field (HDF-S, Williams et al.\ 1999) is a more
recent set of deep {\it HST}\, exposures. The best candidate lens,
noted by the HDF-S team, consists of a blue arc about $0\farcs9$ from
an elliptical galaxy. In \S 2 we report on a detailed study of this
arc. Additional details can be found in Barkana, Blandford, \& Hogg
(1999).

Despite the achievements of {\it HST},\, detecting the
earliest galaxies requires the superior infra-red sensitivity planned
for the Next Generation Space Telescope ({\it NGST}\,; Ferguson
1999). In \S 3 we explore the ability of {\it NGST}\, to extend
gravitational lensing studies well beyond their current limits (see
Barkana \& Loeb 1999 for the complete details). Lensing rates are
expected to increase with source redshift. Sources at $z>10$ will
often be lensed by $z>2$ galaxies, whose masses can then be determined
with lens modeling. In addition, the expected increase by 1--2 orders
of magnitude in the number of extended sources on the sky, due to
observations with {\it NGST},\, will dramatically improve upon the
statistical significance of existing weak lensing measurements.

\section{A possible gravitational lens in the HDF-S}

\begin{figure}
\plotfiddle{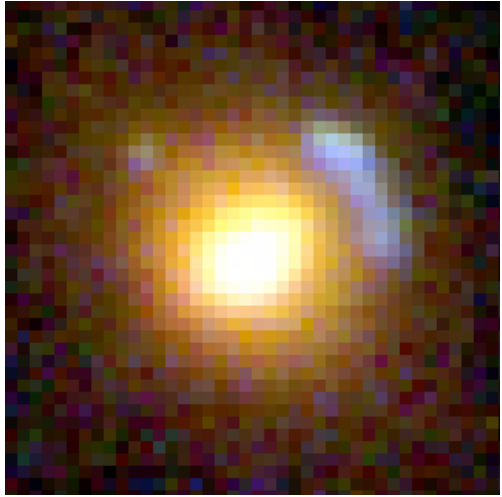}{2.5 in}{0}{58}{58}{-170}{-120}
\caption{A region taken from the public release of the HDF-S 
(Williams et al.\ 1999). It shows the lens galaxy, the arc to its
NW, and the dot to its NE. The figure measures $3\farcs2$ on a side.}
\end{figure}

Figure 1 shows the candidate gravitational lens, HDFS~2232509--603243.
The $V\sim 22$~mag elliptical galaxy fits a de~Vaucouleurs profile
with an effective radius $r_e= 0\farcs31$. Comparison of the colors of
the elliptical with HDF galaxies with known redshifts suggests that
its redshift is roughly $z=0.6$. The $V=25$~mag arc separates
into four distinct components, labeled $A$, $B$, $C$ and $D$, going
clockwise around the arc. We model components $A$, $B$ and $D$
as three images of a common source. We do not model component $C$,
which can be easily produced if the source is extended, or the dot,
which we assume is a separate source.

The image positions can be fit by simple lens models, i.e., a singular
isothermal halo with ellipticity or with shear. Alternatively, a
constant mass-to-light ratio model (based on the resolved image of the
lens galaxy) also provides a fit when combined with a small external
shear. However, with all the models a good fit is obtained only with
a substantial shear produced by objects lying to the NE or SW. There
are several galaxies $\sim 20\arcsec$ to the SW and others $\sim
35\arcsec$ to the NE, and these could contribute the required shear if
spectroscopic redshifts confirm the presence of galaxy groups.

Assuming the photometric lens redshift of $z\sim 0.6$ gives the lens a
luminosity of $\sim L^{\ast}$. Lens models imply a mass-to-light ratio
of $\sim 15$ in solar units, and a velocity dispersion of $\sim
280~{\rm km\ s^{-1}}$. Spectroscopy of this lens system should easily
yield a lens redshift, and may yield a stellar velocity
dispersion and an arc redshift.

\section{Gravitational Lensing with {\it NGST}}

Numerous galaxies and mini-quasars at redshifts $z \ga 5$ will likely
be imaged with {\it NGST}.\, We apply semi-analytic hierarchical
models of galaxy formation to estimate the rate of multiple imaging of
these sources by intervening gravitational lenses. Popular CDM models
for galaxy formation yield a lensing optical depth of $\sim 1\%$ for
sources at $z\sim 10$. The expected slope of the luminosity function
of the early sources (as estimated by Haiman \& Loeb 1998) implies an
additional magnification bias of $\sim 5$, bringing the fraction of
lensed sources at $z=10$ to $\sim 5\%$. Thus, the estimated number of
detected multiply-imaged sources per field of view of {\it NGST}\, is
roughly 5 for $z>10$ quasars, 10 for $z>5$ quasars, 1--15 for $z>10$
galaxies, and 30--200 for $z>5$ galaxies. These ranges in the
numbers of galaxies arise from the uncertain efficiency with which
cooled gas is converted to stars at high redshift. Observations
suggest that this efficiency $\eta$ lies between $2\%$ and $20\%$. 

\begin{figure}
%\plottwo{SBbw.ps}{disk4bw.ps}
\plottwo{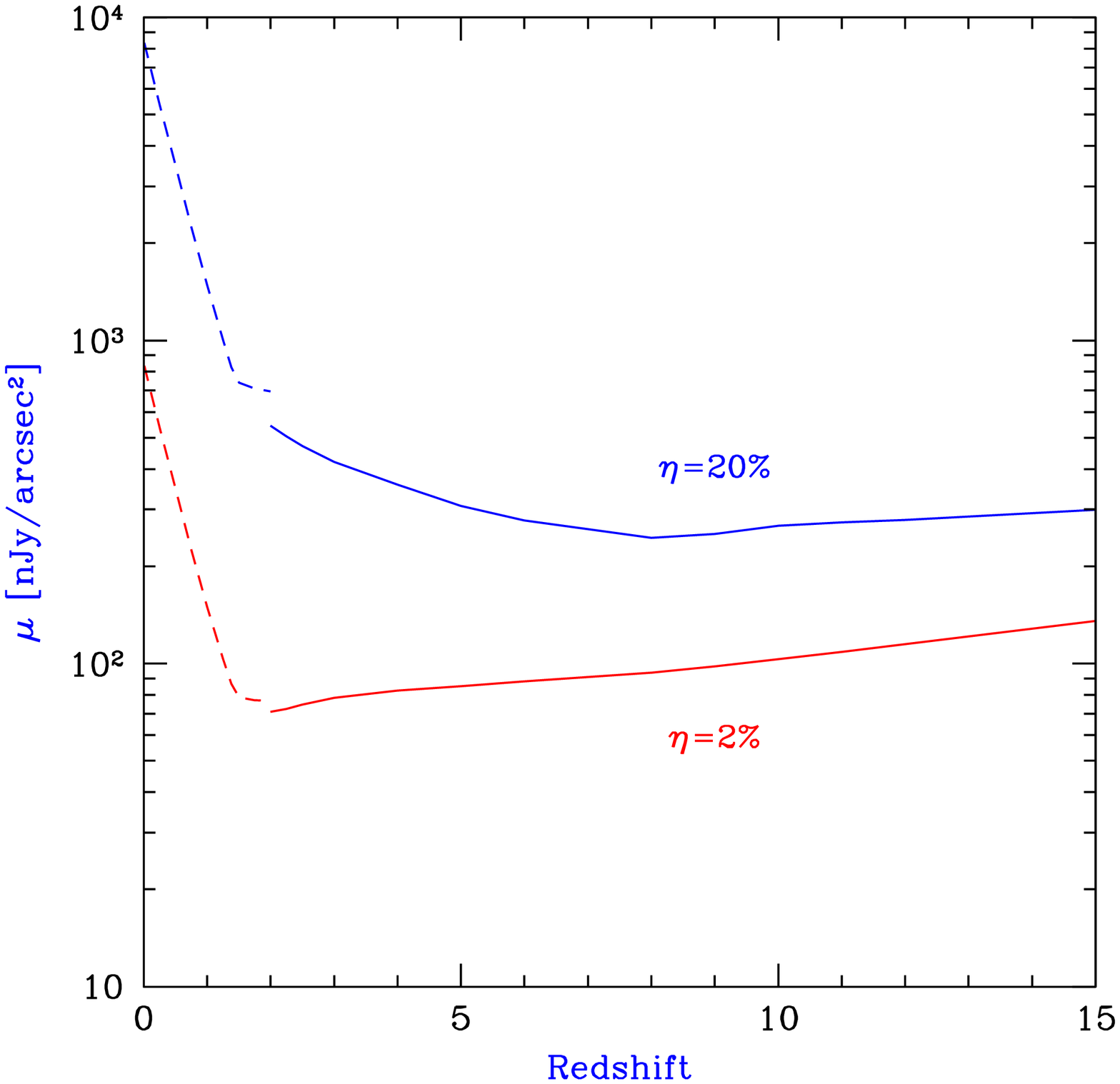}{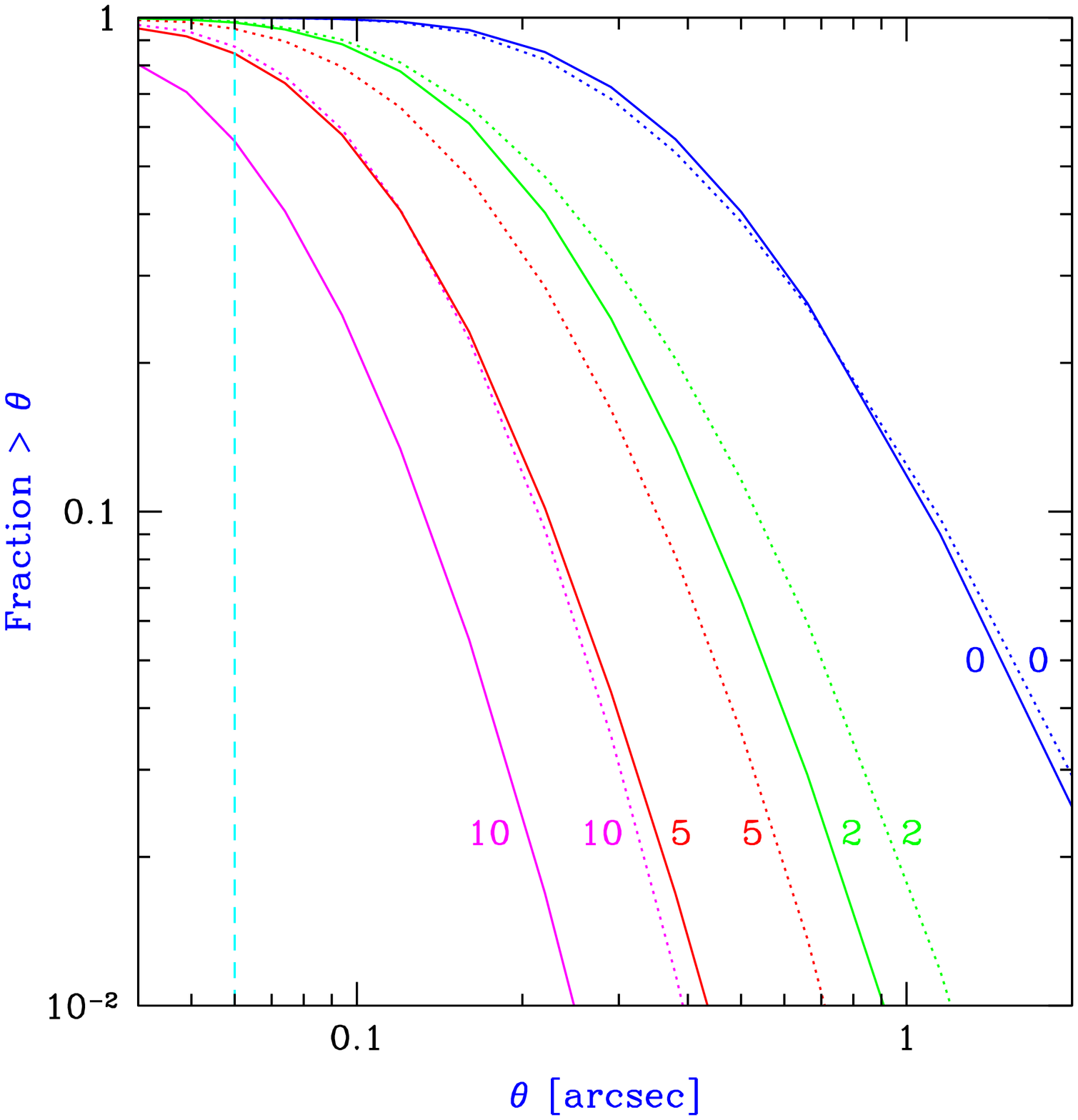}
\caption{The left panel shows the mean surface brightness $\mu$ of
galaxies, averaged over the {\it NGST} wavelength band, for source
galaxies (solid lines) and for lenses (dashed lines). In each case,
the upper and lower curves assume a star formation efficiency
$\eta=20\%$ and $\eta=2\%$, respectively. The curves in the right
panel show the fraction of galaxies with diameters larger than
$\theta$, for $\eta=20\%$ (solid lines) and $\eta=2\%$ (dotted lines).
Different curves consider sources with $0<z<2$, $2<z<5$, $5<z<10$, and
$z>10$, with the lower limit indicated. The {\it NGST}\, resolution
is $0\farcs 06$ (dashed line).}
\end{figure}

Lensed sources may be difficult to detect if their images overlap the
lensing galaxy, and if the lensing galaxy has a higher surface
brightness. In order to compare the surface brightness of source
galaxies to that of lens galaxies, we calculate the redshift evolution
of the mean surface brightness of galaxies detected by {\it NGST},\,
with results shown in Figure 2 (left panel). Although the surface
brightness of a background source will typically be somewhat lower
than that of the foreground lens, the lensed images should be
detectable since they are offset from the lens center and their colors
are expected to differ from those of the lens galaxy. These helpful
features are evident in the case of the HDF-S lens of \S 2.

Weak lensing requires that background sources be resolved.
Figure 2 (right panel) shows the distribution of galaxy sizes at
various redshifts. Although the typical size of sources decreases with
increasing redshift, {\it NGST}\, should resolve at least $60\%$ of
the $z>10$ galaxies that it will detect, and an even larger fraction
of galaxies below redshift 10. This implies that the shapes of these
high redshift galaxies can be studied with {\it NGST},\, and thus the
high resolution of {\it NGST}\, is crucial in making the majority of
sources on the sky useful for weak lensing studies.

In conclusion, the field of gravitational lensing is likely to benefit
greatly over the next decade from the combination of unprecedented
sensitivity and high angular resolution of {\it NGST}.\,

\acknowledgments 

The HDF-S was observed with the NASA/ESA Hubble Space Telescope, which
is operated by AURA under NASA contract NAS 5-26555. Regarding the
predictions for {\it NGST},\, we are grateful to Zoltan Haiman for
number count data, and to Tal Alexander and Amiel Sternberg for
stellar population models. Barkana acknowledges support by Institute
Funds. Loeb acknowledges NASA grants NAG 5-7039 and NAG
5-7768. Blandford acknowledges support by the Alfred P. Sloan
Foundation during a stay at IAS. Hogg acknowledges Hubble Fellowship
grant HF-01093.01-97A from STScI, which is operated by AURA under NASA
contract NAS~5-26555.

\end{document}